\documentclass{caosp309}

\usepackage{graphicx}

\usepackage{natbib}
\usepackage{lipsum}  
\usepackage{ulem}

\articleNo{-}
\pubyear{2024}
\volume{-}
\volnumber{-}
\firstpage{1}
\received{October 2024}
\accepted{?}

\begin{document}

%%%%%%%%%%%%%%%%%%%%%%%%%%%%%%%%%%%%%%%%%%%%%%%%%%%%%%%%%%%%%%%%%%%%%%%%%%%%%
%              R U N N I N G   P A G E   H E A D I N G S                     
% Odd page headings (except for the title page) are produced automatically
% and contain the title. If, and only if, the title of your article is too
% long the running head is omitted in the printout; you can make your own
% running title by using the \htitle command, putting the shortened title
% between the curly brackets. \htitle should also be used when the
% subtitle is present: \htitle offers you a way how to include it into the
% headings. If you wish to see how it works simply remove the % sign from
% the beginning of that line.
%
% Unlike the \htitle command, the \hauthor command is compulsory. It is
% used to produce even page headings and contains the names of the authors
% of an article.  All authors must be listed here, if possible. When
% authors' list is too long, you can abbreviate it by using "{\it et
% al.}". Authors' names are given in the form: initial(s) of the author's
% first name and surname. Authors are separated by a "," (comma) sign and
% the last one by "and".
%%%%%%%%%%%%%%%%%%%%%%%%%%%%%%%%%%%%%%%%%%%%%%%%%%%%%%%%%%%%%%%%%%%%%%%%%%%%%
%\htitle{A note to comet ejection process ...}
\hauthor{D.\,Jones}
%\hauthor{L.\,Neslu\v{s}an {\it et al.}}

%%%%%%%%%%%%%%%%%%%%%%%%%%%%%%%%%%%%%%%%%%%%%%%%%%%%%%%%%%%%%%%%%%%%%%%%%%%%%
%                       T I T L E                                            
% Capital letters in the title are only used at the beginning of the
% names. Don`t end the title by a "." (dot)
%%%%%%%%%%%%%%%%%%%%%%%%%%%%%%%%%%%%%%%%%%%%%%%%%%%%%%%%%%%%%%%%%%%%%%%%%%%%%
\title{Post-common-envelope planetary nebulae}

%%%%%%%%%%%%%%%%%%%%%%%%%%%%%%%%%%%%%%%%%%%%%%%%%%%%%%%%%%%%%%%%%%%%%%%%%%%%%
%                       S U B T I T L E                                      
% You can use the subtitle, with the command \subtitle similar to the
% \title command.
%%%%%%%%%%%%%%%%%%%%%%%%%%%%%%%%%%%%%%%%%%%%%%%%%%%%%%%%%%%%%%%%%%%%%%%%%%%%%

%%%%%%%%%%%%%%%%%%%%%%%%%%%%%%%%%%%%%%%%%%%%%%%%%%%%%%%%%%%%%%%%%%%%%%%%%%%%%
%                   A U T H O R  N A M E S                                   
% Authors' names are separated by the \and command and their institutes
% are assigned by the \inst{n} command. 
% If all authors belong to just one institute, it is not needed/desired
% to use the \inst command.
%
% Author can indicate her/his ORCID (https://orcid.org/) identifier using
% the command \orcid. It will not appear in the LaTeX output but will be
% sent to the ADS database.
%
% When the name contains "Slovak" letters L,d,t,l with a caron, use an
% a new \softl, etc. command (examples given in the last table of
% this document) to produce typographically correct accented characters.
%%%%%%%%%%%%%%%%%%%%%%%%%%%%%%%%%%%%%%%%%%%%%%%%%%%%%%%%%%%%%%%%%%%%%%%%%%%%%
\author{
        D.\,Jones\inst{1,2,3}\orcid{0000-0003-3947-5946}
       }

%%%%%%%%%%%%%%%%%%%%%%%%%%%%%%%%%%%%%%%%%%%%%%%%%%%%%%%%%%%%%%%%%%%%%%%%%%%%%
%           I N S T I T U T E S'  A D D R E S S E S                          
% The affiliation of authors is generated by the \institute command, the
% \and command being again used to separate individual addresses.
% The following commands may be used for the following three institutes:   
%               \lomnica        for      AsU SAV, Tatranska Lomnica          
%               \blava          for      AsU SAV, Bratislava                 
%               \ondrejov       for      AsU CAV, Ondrejov                   
%
% The given postal address must be complete in order to facilitate our
% editorial work. Moreover, you can add your e-mail address, using the
% \email command.
%%%%%%%%%%%%%%%%%%%%%%%%%%%%%%%%%%%%%%%%%%%%%%%%%%%%%%%%%%%%%%%%%%%%%%%%%%%%%
\institute{
           Instituto de Astrof\'isica de Canarias, 38205 La Laguna, Tenerife, Spain, \email{djones@iac.es}
         \and 
           Departamento de Astrof\'isica, Universidad de La Laguna, E-38206 La Laguna, Tenerife, Spain, 
         \and 
           Nordic Optical Telescope, Rambla Jos\'e Ana Fern\'andez P\'erez 7, 38711, Bre\~na Baja, Spain,
          }

%%%%%%%%%%%%%%%%%%%%%%%%%%%%%%%%%%%%%%%%%%%%%%%%%%%%%%%%%%%%%%%%%%%%%%%%%%%%%
%                        D A T E / R E C E I V E D                          
% Date inserted here will be the date when your paper was received The
% format is: month (not abbreviated), day, year.
%%%%%%%%%%%%%%%%%%%%%%%%%%%%%%%%%%%%%%%%%%%%%%%%%%%%%%%%%%%%%%%%%%%%%%%%%%%%%
\date{March 8, 2003}

%%%%%%%%%%%%%%%%%%%%%%%%%%%%%%%%%%%%%%%%%%%%%%%%%%%%%%%%%%%%%%%%%%%%%%%%%%%%%
%                        M A K E T I T L E
% The beginning part (title, author(s), etc.) of your article must be
% closed by the \maketitle command.
%%%%%%%%%%%%%%%%%%%%%%%%%%%%%%%%%%%%%%%%%%%%%%%%%%%%%%%%%%%%%%%%%%%%%%%%%%%%%
\maketitle

%%%%%%%%%%%%%%%%%%%%%%%%%%%%%%%%%%%%%%%%%%%%%%%%%%%%%%%%%%%%%%%%%%%%%%%%%%%%%
%                        A B S T R A C T,  K E Y W O R D S                   
% Here it is shown how to write an abstract.  Keywords should be placed
% within the "abstract" environment using the command \keywords and they
% should be selected from the thesaurus from Astron.  Astrophys.
% Abstracts. They must be separated from each other by -- (two dashes).
%%%%%%%%%%%%%%%%%%%%%%%%%%%%%%%%%%%%%%%%%%%%%%%%%%%%%%%%%%%%%%%%%%%%%%%%%%%%%
\begin{abstract}
Close-binary central stars of planetary nebulae offer a unique tool with which to study the critical and yet poorly understood common-envelope phase of binary stellar evolution. Furthermore, as the nebula itself is thought to comprise the ionised remnant of the ejected common envelope, such planetary nebulae can be used to directly probe the mass, morphology and dynamics of the ejecta. In this review, I summarise our current understanding of the importance of binarity in the formation of planetary nebulae as well as what they may be able to tell us about the common-envelope phase -- including the possible relationships with other post-common-envelope phenomena like stellar mergers, novae and type Ia supernovae.
\end{abstract}

\section{Introduction}

It is now almost half a century since \citet{paczynski76} first proposed the common-envelope (CE) scenario as the most viable pathway towards the formation of close binaries with an evolved component.  Since then, the evidence for the importance of the CE has only grown, however key aspects of the processes at work remain enigmatic \citep{ivanova13}. At the time, \citet{paczynski76} already recognised the importance of planetary nebulae (PNe) in understanding the CE phase, stating \textit{``Observational discovery of a short period binary being a nucleus of a planetary nebula would provide very important support for the evolutionary scenario''}, and the first example was discovered that same year by \citet{bond76}.  Subsequent studies of post-CE planetary nebulae (PNe) have already taught us much about the CE and they continue to represent a key tool with which to shed light on this critical phase of binary evolution \citep{jones20}.

From the perspective of the PN community, the principal interest in CE evolution is the ease with which it can explain the wide range of aspherical morphologies observed in PNe \citep{balick02,garcia-segura18,demarco22}.  However, until recently, there was a reluctance to accept that binarity was the primary shaping mechanism in PNe due to the lack of known binary central stars.  In the early 2000s, only 16 post-CE PNe were known (a couple of which have since been shown to be wide binaries, false associations or non-PNe) leading to the estimate that roughly 10\% of PNe hosted close binary stars \citep{bond00}.  Exploitation of wide-field surveys, like OGLE, Kepler and TESS, has since led to an explosion of new discoveries \citep{miszalski09,jacoby21,aller24}, taking the total now to over 100 known post-CE central stars\footnote{A regularly updated list is maintained by the author at \url{http://drdjones.net/bCSPN}.}.  These studies have placed a hard lower limit on the binary fraction of 20\%, while other probes (such as radial velocities, composite spectra and infrared excesses) have provided much higher estimates \citep[perhaps as high as 80\%;][]{demarco04,douchin15}.

The minimum 20\% post-CE fraction already has one critical implication for our understanding of PNe -- it is likely that some stars which our textbooks would say should produce an observable PN, in fact, do not. Simply put, based on the main sequence binary fraction and period distribution, one would not expect as many as 20\% of all low- and intermediate-mass stars to experience a CE phase, let alone survive one without merging.  The implication being that, if binaries appear over-abundant, then some single stars previously believed to be PN progenitors must not lead to observable PNe.  Clearly, with such a large uncertainty on the true post-CE binary fraction in PNe (20--80\%), it is similarly uncertain just what fraction of these single-star PN progenitors will not lead to PNe.  However, there are indications that it may indeed be significant.  Population synthesis studies have found that perhaps around one fifth of expected PN progenitors due not lead to observable PNe \citep[potentially consistent with only binaries leading to PNe;][]{moe06}, while as many as 50\% of central stars exhibit properties that are inconsistent with single-star evolutionary tracks \citep{weidmann20}.

As a result of conservation of angular momentum during the ejection, the CE is always expected to be preferentially deposited in the orbital plane of the binary thus leading to nebular symmetry axes which lay perpendicular to this plane.  While only nine PNe have been the subject of sufficiently detailed study to constrain both their nebular inclinations and central binary orbital inclinations, the correlation is striking and entirely consistent with the CE shaping each and every one \citep{hillwig16}.  Indeed, the probability of finding such an alignment in these PNe by chance has been shown to be less than one in a million \citep{munday20}.  Nonetheless, there is significant morphological diversity, even among this sample, meaning that, while the CE has clearly played a role in their shaping, the exact processes at work may vary appreciably  \citep{hillwig16}.

Thus, it now seems beyond doubt that CE evolution is a critical aspect of PN formation and evolution, even though the precise details of this are still the subject of active research \citep{jones17,boffin19}. But, perhaps of greater interest for the wider astronomical community, we are now in a position to use post-CE PNe to try and understand the CE phase itself.  The nebular material in these PNe is thought to comprise the ejected CE, offering a unique opportunity to study the ejecta directly (CE ejecta can also be observed in luminous red novae, but these represent ``failed'' CE events, where the stars merged before fully ejecting the donor's envelope). Furthermore, the short-lived nature of PNe ($\tau\sim$30 kyr) means that post-CE central stars of PNe have not yet had time to adjust their chemical and physical properties following the CE ejection.  As such, post-CE PNe are an invaluable tool with which to study the CE phase leading to potentially interesting conclusions for a wide-range of other post-CE phenomena.

\section{Pre-CE mass transfer}

The first step towards using the population of post-CE central stars of PNe to try and understand the CE phase is to constrain the properties of individual post-CE central stars.  Most have been discovered through their photometric variability meaning that only their orbital periods are known.  In order to determine the properties of the stars themselves (masses, temperature, radii), one must obtain follow-up observations that allow for simultaneous modelling of light and radial velocity curves \citep[e.g.,][]{jones22}, and/or spectral analysis using model atmospheres \citep{hillwig17}.

While the number of central stars with well-constrained properties remains relatively small, principally due to the painstaking nature of the required follow-up observations and the complications of the modelling, one can already draw remarkable conclusions.  Where the companion to the nebula progenitor is a main-sequence (MS) star, they are always\footnote{Except the companion to the central star of PN~M~3-1, which was found by \citet{jones19} to be of a normal size but extremely close to filling its Roche lobe (this system will be discussed more in section \ref{sec:novae}).} found to be inflated (often by a factor of $\times$2--3) with respect to isolated MS stars of the same mass \citep{jones15}.  This inflation is almost certainly due to accretion, with further signposts of this accretion coming from the chemical contamination observed in the companion of the central star of The Necklace nebula \citep{miszalski13}.

Accretion during the CE phase is expected to be minimal \citep{sandquist98}, so it is probable that the accretion onto the companion happens before CE.  Further evidence for accretion prior to the CE comes from the kinematical ages of the polar outflows observed in some post-CE PNe.  These polar outflows, thought to be tracers of the mass transfer/accretion, are generally found to predate the ejection of the central nebula (and thus the ejection of the CE) by a few hundred to a few thousand years \citep{corradi11,miszalski11}.  It seems likely that this pre-CE mass transfer might be associated with a period of wind Roche-lobe overflow which could last for several thousands of years as indicated by the  jets observed in Fleming 1 \citep{boffin12}.

It is important to note that there is little to no inflation observed in the MS companions of more evolved post-CE binaries \citep[indeed these systems are excellent probes of the mass-radius relationship of late-type MS stars;][]{parsons18}.  This might, at first, appear puzzling when one takes into account that the thermal timescales for these stars are too long for them to have relaxed back to a normal state.  However, models predict that for intermediate values of accretion efficiency and rate, the response of fully convective stars is to become partially convective \citep{prialnik85}, and it is thus the convective layer formed in the outer envelope which expands rather than the star as a whole.  The relaxation of this convective layer can then occur on a much shorter timescale \citep{jones22}.

\section{Conditions to experience (and survive) a CE}

The characterization of post-CE central stars has revealed a dearth of early type-companions \citep{brown19}.  This is consistent with the idea that  large mass ratios are required to initiate unstable Roche lobe overflow and thus a CE event \citep{pavlovskii15}. The same characterization efforts have also revealed a number of candidates to have experienced the CE while on the red giant branch (RGB).  These stars are analogous to subdwarf B binaries \citep[e.g.,][]{schaffenroth15} and serve as a clear demonstration that RGB CE events can both be survived (without merging) and produce an observable PN \citep{jones22}.

A handful of central stars have been shown to have white dwarf companions, being detected either by radial velocity variability \citep{boffin12} or photometric variability due to tidal deformations \citep{santander15} -- as opposed to the sinusoidal photometric variability due to irradiation which is typical for MS companions.  If one extrapolates from this to assume that all central stars showing photometric variability due to tidal deformations are double degenerates \citep[DD;][]{hillwig10}, then a very significant fraction of all post-CE central stars are DD (particularly when one takes into account that these systems should be harder to detect due to the low amplitudes of variability for most configurations).    These DD post-CE central stars may have formed through two consecutive CE phases \citep{nelemans05}, but their properties are more consistent with having avoided a CE when the first component evolved off the asymptotic giant branch (AGB), with the mass transfer at this stage being stable due to the initial mass ratio being close to unity \citep{woods12}. Understanding the frequency and properties (namely masses and orbital periods) of DD central stars of PNe will likely have important implications for the constraining the pathways towards type \textsc{i}a supernovae \citep{santander15}.

\section{The missing mass problem}

In the single-star scenario, the bright central shells of PNe contain only a small fraction of the mass that  originally comprised the progenitor's envelope, with the vast majority residing in a very low surface brightness halo \citep{villaver02}.  In the case of a CE event, the entirety of the donor's envelope is expected to be ejected on a very short timescale, subsequently forming the PN.  Thus one would expect post-CE PNe to be significantly more massive than their single star counterparts, however \citet{santander22} showed that, in fact, there is very little difference between the masses of the two populations.  Furthermore, studying the observed masses of four PNe for which data of the central stars permits the reconstruction of the CE phase and an estimation of the envelope mass, they found that the masses of the PNe represent only a fraction of the envelope masses of the donors (roughly 60\% for the two DD systems analysed, but as low as 1\% for Abell 63 which has a MS companion).  This brings into question whether the nebulae surrounding post-CE central stars are indeed the ejected CE or rather the product of some much later phase of mass-loss when the CE has already dissipated \citep{corradi15}.

\section{Nebular abundances and the link to novae}
\label{sec:novae}

There are obvious links between post-CE PNe and novae, in particular the discovery of an ancient PN surrounding nova V458 Vul \citep{wesson08}.  However, there are also other less obvious connections.  The central star of PN~M~3-1 was found by \citet{jones19} to have a MS companion which is very close to Roche-lobe filling. This means that as the system evolves there is potential for mass transfer from the companion onto the central star in the relatively near future, possibly leading to a nova eruption inside the expanding PN similar to that of Nova V458 Vul.

Further connections between post-CE PNe and novae are found in their abundance patterns, with some of the shortest period central stars being associated with extreme abundance discrepancy PNe \citep{wesson18}.  The abundances in these PNe are discrepant by a factor of five or more depending on whether they are derived using recombination lines or collisionally-excited emission lines.  These discrepancies indicate the presence of a second (often centrally-concentrated) high-metallicty gas phase in these PNe \citep{garcia16}, where the abundances of this second gas phase are similar to those observed in neon novae \citep{wesson08b}.

\section{Pre-PNe and the link to luminous red novae}

Pre-PNe are thought to represent the intermediate phase between the central star having left the AGB but not yet being hot enough to fully ionise the surrounding nebula. As such, one might expect to find a similar post-CE fraction inside pre-PNe to that found in PNe. However, despite years of dedicated study, no post-CE central star has been found inside a pre-PN \citep[see][and many, many more]{hrivnak11,hrivnak17,hrivnak22,hrivnak24}.  This is perhaps indicative that the CE leads to a shortened pre-PN phase (reducing the likelihood of discovery) or even that pre-PNe and post-CE PNe represent distinct evolutionary pathways.  

Intriguingly, the ejecta properties of luminous red novae -- a class of transients associated with CE mergers -- are strikingly similar to those of pre-PNe \citep{kaminski18}, offering a potential explanation for the dearth of surviving post-CE binaries in pre-PNe.  However, it must be highlighted that much progress has been made linking the morphologies of pre-PNe with wide stellar companions \citep{decin20}, several of which are known in both pre-PNe \citep[e.g.,][]{sanchez04} and PNe \citep[e.g.,][]{jones17b}.

\section{Conclusions}

Binary evolution (both CE and wider) plays a key role in the formation of a significant fraction, perhaps the majority, of PNe (some might even argue all!).  This makes PNe an important test bed with which to study binary stellar evolution -- in particular the poorly understood CE phase, which has been demonstrated to be responsible for at least one fifth of all PNe.  The fact that post-CE PNe should be among the youngest observable post-CE systems,  potentially allowing for direct observations of both the ejected envelope and the surviving post-CE binary, furthermore makes them a unique probe of the CE.

The observed population of post-CE PNe confirms the theoretical expectations that the CE is preferentially ejected in the orbital plane of the binary and that a relatively large initial mass ratio is required in order for unstable mass transfer to occurr.  There are also strong indications of pre-CE mass transfer which can chemically contaminate the companion \citep[now the favoured hypothesis for the formation of carbon dwarfs;][]{whitehouse21} as well as (temporarily) causing their radii to grow. However, not all properties of post-CE PNe are so easily understood -- the so-called missing mass problem, and the origins of a second chemically-enriched gas phase in high abundance discrepancy nebulae are particularly puzzling. The resolution of these puzzles will undoubtedly further our understanding of the CE phase as well as of other CE progeny.

\acknowledgements
DJ would like to thank the organisers of Kopal 2024 for their kind invitation, which was particularly meaningful given that Zden\u{e}k Kopal is his \textit{academic} great-great grandfather (i.e.\ the PhD supervisor of his PhD supervisor's PhD supervisor's PhD supervisor). 

DJ acknowledges support from the Agencia Estatal de Investigaci\'on del Ministerio de Ciencia, Innovaci\'on y Universidades (MCIU/AEI) and the European Regional Development Fund (ERDF) with reference PID-2022-136653NA-I00 (DOI:10.13039/501100011033). DJ also acknowledges support from the Agencia Estatal de Investigaci\'on del Ministerio de Ciencia, Innovaci\'on y Universidades (MCIU/AEI) and the the European Union NextGenerationEU/PRTR with reference CNS2023-143910 (DOI:10.13039/501100011033).

\bibliography{kopal_djones}

\begin{thebibliography}{53}
\expandafter\ifx\csname natexlab\endcsname\relax\def\natexlab#1{#1}\fi

%ADS_ID aller24
\bibitem[{{Aller} {et~al.}(2024){Aller}, {Lillo-Box}, \& {Jones}}]{aller24}
{Aller}, A., {Lillo-Box}, J., \& {Jones}, D., {Planetary nebulae seen with
  TESS: New and revisited short-period binary central star candidates from
  Cycles 1 to 4}. 2024, {\it \aap}, {\bf 690}, A190, DOI:
  10.1051/0004-6361/202450942

%ADS_ID balick02
\bibitem[{{Balick} \& {Frank}(2002)}]{balick02}
{Balick}, B. \& {Frank}, A., {Shapes and Shaping of Planetary Nebulae}. 2002,
  {\it \araa}, {\bf 40}, 439, DOI: 10.1146/annurev.astro.40.060401.093849

%ADS_ID boffin19
\bibitem[{{Boffin} \& {Jones}(2019)}]{boffin19}
{Boffin}, H. M.~J. \& {Jones}, D. 2019, {\it {The Importance of Binaries in the
  Formation and Evolution of Planetary Nebulae}} (Springer Cham)

%ADS_ID boffin12
\bibitem[{{Boffin} {et~al.}(2012){Boffin}, {Miszalski}, {Rauch}, {Jones},
  {Corradi}, {Napiwotzki}, {Day-Jones}, \& {K{\"o}ppen}}]{boffin12}
{Boffin}, H. M.~J., {Miszalski}, B., {Rauch}, T., {et~al.}, {An Interacting
  Binary System Powers Precessing Outflows of an Evolved Star}. 2012, {\it
  Science}, {\bf 338}, 773, DOI: 10.1126/science.1225386

%ADS_ID bond76
\bibitem[{{Bond}(1976)}]{bond76}
{Bond}, H.~E., {Objects common to the Catalogue of Galactic Planetary Nebulae
  and the General Catalogue of Variable Stars.} 1976, {\it \pasp}, {\bf 88},
  192, DOI: 10.1086/129923

%ADS_ID bond00
\bibitem[{{Bond}(2000)}]{bond00}
{Bond}, H.~E., {Binarity of Central Stars of Planetary Nebulae}. 2000, in
  Astronomical Society of the Pacific Conference Series, Vol. {\bf  199}, {\it
  Asymmetrical Planetary Nebulae II: From Origins to Microstructures}, ed.
  J.~H. {Kastner}, N.~{Soker}, \& S.~{Rappaport}, 115

%ADS_ID brown19
\bibitem[{{Brown} {et~al.}(2019){Brown}, {Jones}, {Boffin}, \& {Van
  Winckel}}]{brown19}
{Brown}, A.~J., {Jones}, D., {Boffin}, H. M.~J., \& {Van Winckel}, H., {On the
  post-common-envelope central star of the planetary nebula NGC 2346}. 2019,
  {\it \mnras}, {\bf 482}, 4951, DOI: 10.1093/mnras/sty2986

%ADS_ID corradi15
\bibitem[{{Corradi} {et~al.}(2015){Corradi}, {Garc{\'\i}a-Rojas}, {Jones}, \&
  {Rodr{\'\i}guez-Gil}}]{corradi15}
{Corradi}, R. L.~M., {Garc{\'\i}a-Rojas}, J., {Jones}, D., \&
  {Rodr{\'\i}guez-Gil}, P., {Binarity and the Abundance Discrepancy Problem in
  Planetary Nebulae}. 2015, {\it \apj}, {\bf 803}, 99, DOI:
  10.1088/0004-637X/803/2/99

%ADS_ID corradi11
\bibitem[{{Corradi} {et~al.}(2011){Corradi}, {Sabin}, {Miszalski},
  {Rodr{\'\i}guez-Gil}, {Santander-Garc{\'\i}a}, {Jones}, {Drew}, {Mampaso},
  {Barlow}, {Rubio-D{\'\i}ez}, {Casares}, {Viironen}, {Frew}, {Giammanco},
  {Greimel}, \& {Sale}}]{corradi11}
{Corradi}, R.~L.~M., {Sabin}, L., {Miszalski}, B., {et~al.}, {The Necklace:
  equatorial and polar outflows from the binary central star of the new
  planetary nebula IPHASX J194359.5+170901}. 2011, {\it \mnras}, {\bf 410},
  1349, DOI: 10.1111/j.1365-2966.2010.17523.x

%ADS_ID demarco22
\bibitem[{{De Marco} {et~al.}(2022){De Marco}, {Akashi}, {Akras}, {Alcolea},
  {Aleman}, {Amram}, {Balick}, {De Beck}, {Blackman}, {Boffin}, {Boumis},
  {Bublitz}, {Bucciarelli}, {Bujarrabal}, {Cami}, {Chornay}, {Chu}, {Corradi},
  {Frank}, {Garc{\'\i}a-Hern{\'a}ndez}, {Garc{\'\i}a-Rojas},
  {Garc{\'\i}a-Segura}, {G{\'o}mez-Llanos}, {Gon{\c{c}}alves}, {Guerrero},
  {Jones}, {Karakas}, {Kastner}, {Kwok}, {Lykou}, {Manchado}, {Matsuura},
  {McDonald}, {Miszalski}, {Mohamed}, {Monreal-Ibero}, {Monteiro}, {Montez},
  {Baez}, {Morisset}, {Nordhaus}, {Mendes de Oliveira}, {Osborn}, {Otsuka},
  {Parker}, {Peeters}, {Quint}, {Quintana-Lacaci}, {Redman}, {Ruiter}, {Sabin},
  {Sahai}, {Contreras}, {Santander-Garc{\'\i}a}, {Seitenzahl}, {Soker},
  {Speck}, {Stanghellini}, {Steffen}, {Toal{\'a}}, {Ueta}, {Van de Steene},
  {Van Winckel}, {Ventura}, {Villaver}, {Vlemmings}, {Walsh}, {Wesson}, \&
  {Zijlstra}}]{demarco22}
{De Marco}, O., {Akashi}, M., {Akras}, S., {et~al.}, {The messy death of a
  multiple star system and the resulting planetary nebula as observed by JWST}.
  2022, {\it Nature Astronomy}, {\bf 6}, 1421, DOI: 10.1038/s41550-022-01845-2

%ADS_ID demarco04
\bibitem[{{De Marco} {et~al.}(2004){De Marco}, {Bond}, {Harmer}, \&
  {Fleming}}]{demarco04}
{De Marco}, O., {Bond}, H.~E., {Harmer}, D., \& {Fleming}, A.~J., {Indications
  of a Large Fraction of Spectroscopic Binaries among Nuclei of Planetary
  Nebulae}. 2004, {\it \apjl}, {\bf 602}, L93, DOI: 10.1086/382156

%ADS_ID decin20
\bibitem[{{Decin} {et~al.}(2020){Decin}, {Montarg{\`e}s}, {Richards},
  {Gottlieb}, {Homan}, {McDonald}, {El Mellah}, {Danilovich}, {Wallstr{\"o}m},
  {Zijlstra}, {Baudry}, {Bolte}, {Cannon}, {De Beck}, {De Ceuster}, {de Koter},
  {De Ridder}, {Etoka}, {Gobrecht}, {Gray}, {Herpin}, {Jeste}, {Lagadec},
  {Kervella}, {Khouri}, {Menten}, {Millar}, {M{\"u}ller}, {Plane}, {Sahai},
  {Sana}, {Van de Sande}, {Waters}, {Wong}, \& {Yates}}]{decin20}
{Decin}, L., {Montarg{\`e}s}, M., {Richards}, A.~M.~S., {et~al.}, {(Sub)stellar
  companions shape the winds of evolved stars}. 2020, {\it Science}, {\bf 369},
  1497, DOI: 10.1126/science.abb1229

%ADS_ID douchin15
\bibitem[{{Douchin} {et~al.}(2015){Douchin}, {De Marco}, {Frew}, {Jacoby},
  {Jasniewicz}, {Fitzgerald}, {Passy}, {Harmer}, {Hillwig}, \&
  {Moe}}]{douchin15}
{Douchin}, D., {De Marco}, O., {Frew}, D.~J., {et~al.}, {The binary fraction of
  planetary nebula central stars - II. A larger sample and improved technique
  for the infrared excess search}. 2015, {\it \mnras}, {\bf 448}, 3132, DOI:
  10.1093/mnras/stu2700

%ADS_ID garcia16
\bibitem[{{Garc{\'\i}a-Rojas} {et~al.}(2016){Garc{\'\i}a-Rojas}, {Corradi},
  {Monteiro}, {Jones}, {Rodr{\'\i}guez-Gil}, \& {Cabrera-Lavers}}]{garcia16}
{Garc{\'\i}a-Rojas}, J., {Corradi}, R. L.~M., {Monteiro}, H., {et~al.},
  {Imaging the Elusive H-poor Gas in the High adf Planetary Nebula NGC 6778}.
  2016, {\it \apjl}, {\bf 824}, L27, DOI: 10.3847/2041-8205/824/2/L27

%ADS_ID garcia-segura18
\bibitem[{{Garc{\'\i}a-Segura} {et~al.}(2018){Garc{\'\i}a-Segura}, {Ricker}, \&
  {Taam}}]{garcia-segura18}
{Garc{\'\i}a-Segura}, G., {Ricker}, P.~M., \& {Taam}, R.~E., {Common Envelope
  Shaping of Planetary Nebulae}. 2018, {\it \apj}, {\bf 860}, 19, DOI:
  10.3847/1538-4357/aac08c

%ADS_ID hillwig10
\bibitem[{{Hillwig} {et~al.}(2010){Hillwig}, {Bond}, {Af{\c{s}}ar}, \& {De
  Marco}}]{hillwig10}
{Hillwig}, T.~C., {Bond}, H.~E., {Af{\c{s}}ar}, M., \& {De Marco}, O., {Binary
  Central Stars of Planetary Nebulae Discovered Through Photometric
  Variability. II. Modeling the Central Stars of NGC 6026 and NGC 6337}. 2010,
  {\it \aj}, {\bf 140}, 319, DOI: 10.1088/0004-6256/140/2/319

%ADS_ID hillwig17
\bibitem[{{Hillwig} {et~al.}(2017){Hillwig}, {Frew}, {Reindl}, {Rotter},
  {Webb}, \& {Margheim}}]{hillwig17}
{Hillwig}, T.~C., {Frew}, D.~J., {Reindl}, N., {et~al.}, {Binary Central Stars
  of Planetary Nebulae Discovered Through Photometric Variability. V. The
  Central Stars of HaTr 7 and ESO 330-9}. 2017, {\it \aj}, {\bf 153}, 24, DOI:
  10.3847/1538-3881/153/1/24

%ADS_ID hillwig16
\bibitem[{{Hillwig} {et~al.}(2016){Hillwig}, {Jones}, {De Marco}, {Bond},
  {Margheim}, \& {Frew}}]{hillwig16}
{Hillwig}, T.~C., {Jones}, D., {De Marco}, O., {et~al.}, {Observational
  Confirmation of a Link Between Common Envelope Binary Interaction and
  Planetary Nebula Shaping}. 2016, {\it \apj}, {\bf 832}, 125, DOI:
  10.3847/0004-637X/832/2/125

%ADS_ID hrivnak22
\bibitem[{{Hrivnak} {et~al.}(2022){Hrivnak}, {Lu}, {Bakke}, \&
  {Grimm}}]{hrivnak22}
{Hrivnak}, B.~J., {Lu}, W., {Bakke}, W.~C., \& {Grimm}, P.~J., {Variability in
  Protoplanetary Nebulae. IX. Evidence for Evolution in a Decade}. 2022, {\it
  \apj}, {\bf 939}, 32, DOI: 10.3847/1538-4357/ac938a

%ADS_ID hrivnak11
\bibitem[{{Hrivnak} {et~al.}(2011){Hrivnak}, {Lu}, {Bohlender}, {Morris},
  {Woodsworth}, \& {Scarfe}}]{hrivnak11}
{Hrivnak}, B.~J., {Lu}, W., {Bohlender}, D., {et~al.}, {Are Proto-planetary
  Nebulae Shaped by a Binary? Results of a Long-term Radial Velocity Study}.
  2011, {\it \apj}, {\bf 734}, 25, DOI: 10.1088/0004-637X/734/1/25

%ADS_ID hrivnak24
\bibitem[{{Hrivnak} {et~al.}(2024){Hrivnak}, {Lu}, {Henson}, \&
  {Hillwig}}]{hrivnak24}
{Hrivnak}, B.~J., {Lu}, W., {Henson}, G., \& {Hillwig}, T.~C., {Variability in
  Protoplanetary Nebulae. X. Multiyear Periods as an Indicator of Potential
  Binaries}. 2024, {\it \aj}, {\bf 167}, 30, DOI: 10.3847/1538-3881/ad0cc4

%ADS_ID hrivnak17
\bibitem[{{Hrivnak} {et~al.}(2017){Hrivnak}, {Van de Steene}, {Van Winckel},
  {Sperauskas}, {Bohlender}, \& {Lu}}]{hrivnak17}
{Hrivnak}, B.~J., {Van de Steene}, G., {Van Winckel}, H., {et~al.}, {Where are
  the Binaries? Results of a Long-term Search for Radial Velocity Binaries in
  Proto-planetary Nebulae}. 2017, {\it \apj}, {\bf 846}, 96, DOI:
  10.3847/1538-4357/aa84ae

%ADS_ID ivanova13
\bibitem[{{Ivanova} {et~al.}(2013){Ivanova}, {Justham}, {Chen}, {De Marco},
  {Fryer}, {Gaburov}, {Ge}, {Glebbeek}, {Han}, {Li}, {Lu}, {Marsh},
  {Podsiadlowski}, {Potter}, {Soker}, {Taam}, {Tauris}, {van den Heuvel}, \&
  {Webbink}}]{ivanova13}
{Ivanova}, N., {Justham}, S., {Chen}, X., {et~al.}, {Common envelope evolution:
  where we stand and how we can move forward}. 2013, {\it \aapr}, {\bf 21}, 59,
  DOI: 10.1007/s00159-013-0059-2

%ADS_ID jacoby21
\bibitem[{{Jacoby} {et~al.}(2021){Jacoby}, {Hillwig}, {Jones}, {Martin}, {De
  Marco}, {Kronberger}, {Hurowitz}, {Crocker}, \& {Dey}}]{jacoby21}
{Jacoby}, G.~H., {Hillwig}, T.~C., {Jones}, D., {et~al.}, {Binary central stars
  of planetary nebulae identified with Kepler/K2}. 2021, {\it \mnras}, {\bf
  506}, 5223, DOI: 10.1093/mnras/stab2045

%ADS_ID jones20
\bibitem[{{Jones}(2020)}]{jones20}
{Jones}, D., {Observational Constraints on the Common Envelope Phase}. 2020, in
  {\it Reviews in Frontiers of Modern Astrophysics; From Space Debris to
  Cosmology}, ed. P.~{Kab{\'a}th}, D.~{Jones}, \& M.~{Skarka} (Springer),
  123--153

%ADS_ID jones17
\bibitem[{{Jones} \& {Boffin}(2017)}]{jones17}
{Jones}, D. \& {Boffin}, H. M.~J., {Binary stars as the key to understanding
  planetary nebulae}. 2017, {\it Nature Astronomy}, {\bf 1}, 0117, DOI:
  10.1038/s41550-017-0117

%ADS_ID jones15
\bibitem[{{Jones} {et~al.}(2015){Jones}, {Boffin}, {Rodr{\'\i}guez-Gil},
  {Wesson}, {Corradi}, {Miszalski}, \& {Mohamed}}]{jones15}
{Jones}, D., {Boffin}, H.~M.~J., {Rodr{\'\i}guez-Gil}, P., {et~al.}, {The
  post-common envelope central stars of the planetary nebulae Henize 2-155 and
  Henize 2-161}. 2015, {\it \aap}, {\bf 580}, A19, DOI:
  10.1051/0004-6361/201425454

%ADS_ID jones19
\bibitem[{{Jones} {et~al.}(2019){Jones}, {Boffin}, {Sowicka}, {Miszalski},
  {Rodr{\'\i}guez-Gil}, {Santander-Garc{\'\i}a}, \& {Corradi}}]{jones19}
{Jones}, D., {Boffin}, H. M.~J., {Sowicka}, P., {et~al.}, {The short orbital
  period binary star at the heart of the planetary nebula M 3-1}. 2019, {\it
  \mnras}, {\bf 482}, L75, DOI: 10.1093/mnrasl/sly142

%ADS_ID jones22
\bibitem[{{Jones} {et~al.}(2022){Jones}, {Munday}, {Corradi},
  {Rodr{\'\i}guez-Gil}, {Boffin}, {Zak}, {Sowicka}, {Parsons}, {Dhillon},
  {Littlefair}, {Marsh}, {Reindl}, \& {Garc{\'\i}a-Rojas}}]{jones22}
{Jones}, D., {Munday}, J., {Corradi}, R. L.~M., {et~al.}, {The
  post-common-envelope binary central star of the planetary nebula Ou 5: a
  doubly eclipsing post-red-giant-branch system}. 2022, {\it \mnras}, {\bf
  510}, 3102, DOI: 10.1093/mnras/stab3736

%ADS_ID jones17b
\bibitem[{{Jones} {et~al.}(2017){Jones}, {Van Winckel}, {Aller}, {Exter}, \&
  {De Marco}}]{jones17b}
{Jones}, D., {Van Winckel}, H., {Aller}, A., {Exter}, K., \& {De Marco}, O.,
  {The long-period binary central stars of the planetary nebulae NGC 1514 and
  LoTr 5}. 2017, {\it \aap}, {\bf 600}, L9, DOI: 10.1051/0004-6361/201730700

%ADS_ID kaminski18
\bibitem[{{Kami{\'n}ski} {et~al.}(2018){Kami{\'n}ski}, {Steffen}, {Tylenda},
  {Young}, {Patel}, \& {Menten}}]{kaminski18}
{Kami{\'n}ski}, T., {Steffen}, W., {Tylenda}, R., {et~al.}, {Submillimeter-wave
  emission of three Galactic red novae: cool molecular outflows produced by
  stellar mergers}. 2018, {\it \aap}, {\bf 617}, A129, DOI:
  10.1051/0004-6361/201833165

%ADS_ID miszalski09
\bibitem[{{Miszalski} {et~al.}(2009){Miszalski}, {Acker}, {Moffat}, {Parker},
  \& {Udalski}}]{miszalski09}
{Miszalski}, B., {Acker}, A., {Moffat}, A.~F.~J., {Parker}, Q.~A., \&
  {Udalski}, A., {Binary planetary nebulae nuclei towards the Galactic bulge.
  I. Sample discovery, period distribution, and binary fraction}. 2009, {\it
  \aap}, {\bf 496}, 813, DOI: 10.1051/0004-6361/200811380

%ADS_ID miszalski13
\bibitem[{{Miszalski} {et~al.}(2013){Miszalski}, {Boffin}, \&
  {Corradi}}]{miszalski13}
{Miszalski}, B., {Boffin}, H. M.~J., \& {Corradi}, R. L.~M., {A carbon dwarf
  wearing a Necklace: first proof of accretion in a post-common-envelope binary
  central star of a planetary nebula with jets}. 2013, {\it \mnras}, {\bf 428},
  L39, DOI: 10.1093/mnrasl/sls011

%ADS_ID miszalski11
\bibitem[{{Miszalski} {et~al.}(2011){Miszalski}, {Corradi}, {Boffin}, {Jones},
  {Sabin}, {Santander-Garc{\'\i}a}, {Rodr{\'\i}guez-Gil}, \&
  {Rubio-D{\'\i}ez}}]{miszalski11}
{Miszalski}, B., {Corradi}, R.~L.~M., {Boffin}, H.~M.~J., {et~al.}, {ETHOS 1: a
  high-latitude planetary nebula with jets forged by a post-common-envelope
  binary central star}. 2011, {\it \mnras}, {\bf 413}, 1264, DOI:
  10.1111/j.1365-2966.2011.18212.x

%ADS_ID moe06
\bibitem[{{Moe} \& {De Marco}(2006)}]{moe06}
{Moe}, M. \& {De Marco}, O., {Do Most Planetary Nebulae Derive from Binaries?
  I. Population Synthesis Model of the Galactic Planetary Nebula Population
  Produced by Single Stars and Binaries}. 2006, {\it \apj}, {\bf 650}, 916,
  DOI: 10.1086/506900

%ADS_ID munday20
\bibitem[{{Munday} {et~al.}(2020){Munday}, {Jones}, {Garc{\'\i}a-Rojas},
  {Boffin}, {Miszalski}, {Corradi}, {Rodr{\'\i}guez-Gil}, {Rubio-D{\'\i}ez},
  {Santander-Garc{\'\i}a}, \& {Sowicka}}]{munday20}
{Munday}, J., {Jones}, D., {Garc{\'\i}a-Rojas}, J., {et~al.}, {The
  post-common-envelope binary central star of the planetary nebula ETHOS 1}.
  2020, {\it \mnras}, {\bf 498}, 6005, DOI: 10.1093/mnras/staa2753

%ADS_ID nelemans05
\bibitem[{{Nelemans} \& {Tout}(2005)}]{nelemans05}
{Nelemans}, G. \& {Tout}, C.~A., {Reconstructing the evolution of white dwarf
  binaries: further evidence for an alternative algorithm for the outcome of
  the common-envelope phase in close binaries}. 2005, {\it \mnras}, {\bf 356},
  753, DOI: 10.1111/j.1365-2966.2004.08496.x

%ADS_ID paczynski76
\bibitem[{{Paczynski}(1976)}]{paczynski76}
{Paczynski}, B., {Common Envelope Binaries}. 1976, in IAU Symposium, Vol. {\bf
  ~73}, {\it Structure and Evolution of Close Binary Systems}, ed.
  P.~{Eggleton}, S.~{Mitton}, \& J.~{Whelan}, 75

%ADS_ID parsons18
\bibitem[{{Parsons} {et~al.}(2018){Parsons}, {G{\"a}nsicke}, {Marsh}, {Ashley},
  {Breedt}, {Burleigh}, {Copperwheat}, {Dhillon}, {Green}, {Hermes}, {Irawati},
  {Kerry}, {Littlefair}, {Rebassa-Mansergas}, {Sahman}, {Schreiber}, \&
  {Zorotovic}}]{parsons18}
{Parsons}, S.~G., {G{\"a}nsicke}, B.~T., {Marsh}, T.~R., {et~al.}, {The scatter
  of the M dwarf mass-radius relationship}. 2018, {\it \mnras}, {\bf 481},
  1083, DOI: 10.1093/mnras/sty2345

%ADS_ID pavlovskii15
\bibitem[{{Pavlovskii} \& {Ivanova}(2015)}]{pavlovskii15}
{Pavlovskii}, K. \& {Ivanova}, N., {Mass transfer from giant donors}. 2015,
  {\it \mnras}, {\bf 449}, 4415, DOI: 10.1093/mnras/stv619

%ADS_ID prialnik85
\bibitem[{{Prialnik} \& {Livio}(1985)}]{prialnik85}
{Prialnik}, D. \& {Livio}, M., {The outcome of accretion on to a fully
  convective star Expansion or contraction?} 1985, {\it \mnras}, {\bf 216}, 37,
  DOI: 10.1093/mnras/216.1.37

%ADS_ID sanchez04
\bibitem[{{S{\'a}nchez Contreras} {et~al.}(2004){S{\'a}nchez Contreras}, {Gil
  de Paz}, \& {Sahai}}]{sanchez04}
{S{\'a}nchez Contreras}, C., {Gil de Paz}, A., \& {Sahai}, R., {The Companion
  to the Central Mira Star of the Protoplanetary Nebula OH 231.8+4.2}. 2004,
  {\it \apj}, {\bf 616}, 519, DOI: 10.1086/424827

%ADS_ID sandquist98
\bibitem[{{Sandquist} {et~al.}(1998){Sandquist}, {Taam}, {Chen}, {Bodenheimer},
  \& {Burkert}}]{sandquist98}
{Sandquist}, E.~L., {Taam}, R.~E., {Chen}, X., {Bodenheimer}, P., \& {Burkert},
  A., {Double Core Evolution. X. Through the Envelope Ejection Phase}. 1998,
  {\it \apj}, {\bf 500}, 909, DOI: 10.1086/305778

%ADS_ID santander22
\bibitem[{{Santander-Garc{\'\i}a} {et~al.}(2022){Santander-Garc{\'\i}a},
  {Jones}, {Alcolea}, {Bujarrabal}, \& {Wesson}}]{santander22}
{Santander-Garc{\'\i}a}, M., {Jones}, D., {Alcolea}, J., {Bujarrabal}, V., \&
  {Wesson}, R., {The ionised and molecular mass of post-common-envelope
  planetary nebulae. The missing mass problem}. 2022, {\it \aap}, {\bf 658},
  A17, DOI: 10.1051/0004-6361/202142233

%ADS_ID santander15
\bibitem[{{Santander-Garc{\'\i}a} {et~al.}(2015){Santander-Garc{\'\i}a},
  {Rodr{\'\i}guez-Gil}, {Corradi}, {Jones}, {Miszalski}, {Boffin},
  {Rubio-D{\'\i}ez}, \& {Kotze}}]{santander15}
{Santander-Garc{\'\i}a}, M., {Rodr{\'\i}guez-Gil}, P., {Corradi}, R.~L.~M.,
  {et~al.}, {The double-degenerate, super-Chandrasekhar nucleus of the
  planetary nebula Henize 2-428}. 2015, {\it \nat}, {\bf 519}, 63, DOI:
  10.1038/nature14124

%ADS_ID schaffenroth15
\bibitem[{{Schaffenroth} {et~al.}(2015){Schaffenroth}, {Barlow}, {Drechsel}, \&
  {Dunlap}}]{schaffenroth15}
{Schaffenroth}, V., {Barlow}, B.~N., {Drechsel}, H., \& {Dunlap}, B.~H., {An
  eclipsing post common-envelope system consisting of a pulsating hot subdwarf
  B star and a brown dwarf companion}. 2015, {\it \aap}, {\bf 576}, A123, DOI:
  10.1051/0004-6361/201525701

%ADS_ID villaver02
\bibitem[{{Villaver} {et~al.}(2002){Villaver}, {Manchado}, \&
  {Garc{\'\i}a-Segura}}]{villaver02}
{Villaver}, E., {Manchado}, A., \& {Garc{\'\i}a-Segura}, G., {The Dynamical
  Evolution of the Circumstellar Gas around Low- and Intermediate-Mass Stars.
  II. The Planetary Nebula Formation}. 2002, {\it \apj}, {\bf 581}, 1204, DOI:
  10.1086/344250

%ADS_ID weidmann20
\bibitem[{{Weidmann} {et~al.}(2020){Weidmann}, {Mari}, {Schmidt}, {Gaspar},
  {Miller Bertolami}, {Oio}, {Guti{\'e}rrez-Soto}, {Volpe}, {Gamen}, \&
  {Mast}}]{weidmann20}
{Weidmann}, W.~A., {Mari}, M.~B., {Schmidt}, E.~O., {et~al.}, {Catalogue of the
  central stars of planetary nebulae. Expanded edition}. 2020, {\it \aap}, {\bf
  640}, A10, DOI: 10.1051/0004-6361/202037998

%ADS_ID wesson08
\bibitem[{{Wesson} {et~al.}(2008{\natexlab{a}}){Wesson}, {Barlow}, {Corradi},
  {Drew}, {Groot}, {Knigge}, {Steeghs}, {Gaensicke}, {Napiwotzki},
  {Rodriguez-Gil}, {Zijlstra}, {Bode}, {Drake}, {Frew}, {Gonzalez-Solares},
  {Greimel}, {Irwin}, {Morales-Rueda}, {Nelemans}, {Parker}, {Sale},
  {Sokoloski}, {Somero}, {Uthas}, {Walton}, {Warner}, {Watson}, \&
  {Wright}}]{wesson08}
{Wesson}, R., {Barlow}, M.~J., {Corradi}, R.~L.~M., {et~al.}, {A Planetary
  Nebula around Nova V458 Vulpeculae Undergoing Flash Ionization}.
  2008{\natexlab{a}}, {\it \apjl}, {\bf 688}, L21, DOI: 10.1086/594366

%ADS_ID wesson08b
\bibitem[{{Wesson} {et~al.}(2008{\natexlab{b}}){Wesson}, {Barlow}, {Liu},
  {Storey}, {Ercolano}, \& {De Marco}}]{wesson08b}
{Wesson}, R., {Barlow}, M.~J., {Liu}, X.~W., {et~al.}, {The hydrogen-deficient
  knot of the `born-again' planetary nebula Abell 58 (V605 Aql)}.
  2008{\natexlab{b}}, {\it \mnras}, {\bf 383}, 1639, DOI:
  10.1111/j.1365-2966.2007.12683.x

%ADS_ID wesson18
\bibitem[{{Wesson} {et~al.}(2018){Wesson}, {Jones}, {Garc{\'\i}a-Rojas},
  {Boffin}, \& {Corradi}}]{wesson18}
{Wesson}, R., {Jones}, D., {Garc{\'\i}a-Rojas}, J., {Boffin}, H.~M.~J., \&
  {Corradi}, R.~L.~M., {Confirmation of the link between central star binarity
  and extreme abundance discrepancy factors in planetary nebulae}. 2018, {\it
  \mnras}, {\bf 480}, 4589, DOI: 10.1093/mnras/sty1871

%ADS_ID whitehouse21
\bibitem[{{Whitehouse} {et~al.}(2021){Whitehouse}, {Farihi}, {Howarth},
  {Mancino}, {Walters}, {Swan}, {Wilson}, \& {Guo}}]{whitehouse21}
{Whitehouse}, L.~J., {Farihi}, J., {Howarth}, I.~D., {et~al.}, {Carbon-enhanced
  stars with short orbital and spin periods}. 2021, {\it \mnras}, {\bf 506},
  4877, DOI: 10.1093/mnras/stab1913

%ADS_ID woods12
\bibitem[{{Woods} {et~al.}(2012){Woods}, {Ivanova}, {van der Sluys}, \&
  {Chaichenets}}]{woods12}
{Woods}, T.~E., {Ivanova}, N., {van der Sluys}, M.~V., \& {Chaichenets}, S.,
  {On the Formation of Double White Dwarfs through Stable Mass Transfer and a
  Common Envelope}. 2012, {\it \apj}, {\bf 744}, 12, DOI:
  10.1088/0004-637X/744/1/12

\end{thebibliography}
\end{document}